\documentclass[conference]{IEEEtran}
\usepackage[utf8]{inputenc}
\IEEEoverridecommandlockouts
\usepackage{url}

\usepackage[cmex10]{amsmath} 

\def\BibTeX{{\rm B\kern-.05em{\sc i\kern-.025em b}\kern-.08em
    T\kern-.1667em\lower.7ex\hbox{E}\kern-.125emX}}

\def\proofoutline{\noindent{\it Proof Outline}. \ignorespaces}

\usepackage{cite}
\usepackage{amsmath,amssymb,amsfonts}
\usepackage{algorithmic}
\usepackage{graphicx}
\usepackage{textcomp}
\usepackage{xcolor}
\usepackage{stmaryrd}
\usepackage{mathrsfs}
\usepackage{epic,eepic}
\usepackage{theorem}
\usepackage{algorithm}
\usepackage{mathtools}
\usepackage{amsfonts, dsfont}
\usepackage{cite}
\usepackage{tikz}
\usepackage{pgfplots}
\usetikzlibrary{arrows, arrows.meta, calc}
\usetikzlibrary{patterns}
\usepackage{caption}
\usepackage{enumitem}

\newtheorem{lemma}{Lemma}
\newtheorem{theorem}{Theorem}[section]
\newtheorem{definition}[theorem]{Definition}
\newtheorem{corollary}[theorem]{Corollary}
\newtheorem{proposition}[theorem]{Proposition}

\newtheorem{remark}[theorem]{Remark}
\newtheorem{example}[theorem]{Example}

\newcommand{\N}{\mathbb{N}}

\DeclareMathOperator*{\argmin}{\arg\!\min}
\DeclareMathOperator*{\supp}{supp}

\tikzstyle myBG=[line width=3pt,opacity=1]
\tikzstyle{suite}=[->,>=stealth,very thick]
\usepackage{tikz-cd}
\usetikzlibrary{arrows,positioning}
\tikzset{
  shift left/.style ={commutative diagrams/shift left={#1}},
  shift right/.style={commutative diagrams/shift right={#1}}
}
\title{Optimal Zero-Error Coding for Computing under Pairwise Shared Side Information}
\author{\IEEEauthorblockN{Nicolas Charpenay}
\IEEEauthorblockA{\textit{Institut de Recherche en Informatique}\\
\textit{et Systèmes Aléatoires (IRISA)}\\
Rennes, FRANCE \\
nicolas.charpenay@irisa.fr}
\and
\IEEEauthorblockN{Ma\"el le Treust}
\IEEEauthorblockA{\textit{Institut de Recherche en Informatique}\\
\textit{et Systèmes Aléatoires (IRISA)} \\
Rennes, FRANCE \\
mael.le-treust@irisa.fr}
\and
\IEEEauthorblockN{Aline Roumy}
\IEEEauthorblockA{\textit{Institut National de Recherche} \\
\textit{en Informatique et en Automatique (INRIA)}\\
Rennes, FRANCE \\
aline.roumy@inria.fr}
}

\begin{document}
\maketitle

\begin{abstract}
    We study the zero-error source coding problem in which an encoder with Side Information (SI) $g(Y)$ transmits source symbols $X$ to a decoder. The decoder has SI $Y$ and wants to recover $f(X,Y)$ where $f,g$ are deterministic. We exhibit a condition on the source distribution and $g$ that we call ``pairwise shared side information'', such that the optimal rate has a single-letter expression. This condition is satisfied if every pair of source symbols ``share" at least one SI symbol for all output of $g$. It has a practical interpretation, as $Y$ models a request made by the encoder on an image $X$, and $g(Y)$ corresponds to the type of request. It also has a graph-theoretical interpretation: under ``pairwise shared side information'' the characteristic graph can be written as a disjoint union of OR products. In the case where the source distribution is full-support, we provide an analytic expression for the optimal rate. We develop an example under ``pairwise shared side information'', and we show that the optimal coding scheme outperforms several strategies from the literature.
\end{abstract}


\section{Introduction}

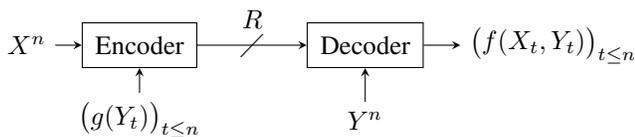
\begin{figure}[b!]
    \centering
    \begin{tikzpicture}
    \node[shape=rectangle, draw=black, fill=white, inner sep=5pt] (E) at (0,0) {Encoder};
    \node[shape=rectangle, draw=black, fill=white, inner sep=5pt] (D) at (3,0) {Decoder};
    
    \node[draw=none] (SZ) at ($(E)+(0,-1)$) {$\big(g(Y_t)\big)_{t \leq n}$};
    \node[draw=none] (SY) at ($(D)+(0,-1)$) {$Y^n$};
    \node[draw=none] (X_1) at ($(D)+(2.5,0)$) {$\big(f(X_t, Y_t)\big)_{t \leq n}$};
    \node[draw=none] (X) at ($(E)+(-1.5,0)$) {$X^n$};
    
    \draw[->, >=stealth] (D) edge (X_1);
    \draw[->, >=stealth] (X) edge (E);
    \draw[->, >=stealth] (SY) edge (D);
    \draw[->, >=stealth] (SZ) edge (E);
    \draw[->, >=stealth] (E) edge (D);
    
    \node[draw=none] (R0) at ($(1.5,0)$) {$\diagup$};
    \node[draw=none] (R) at ($(R0)+(0,10pt)$) {$R$};
    
    \end{tikzpicture}
    \caption{Zero-error coding for computing with side information at the encoder.}
    \label{fig:Setting2}
\end{figure}

Consider the source coding scenario depicted in Figure \ref{fig:Setting2} where two correlated sequences $(X^n,Y^n)$ of discrete i.i.d. random source symbols are drawn with a distribution $P_{X,Y}$. The encoder knows $X^n$, has $(g(Y_t))_{t \leq n}$ as side information, and transmits information to the decoder through a perfect channel of capacity $R$. The decoder has the side information $Y^n$ and wants to reconstruct $(f(X_t, Y_t))_{t \leq n}$, where $f,g$ are deterministic. What is the minimal $R$ such that $(f(X_t, Y_t))_{t \leq n}$ can be retrieved by the decoder with probability of error $0$?

This coding problem appears in video compression \cite{duan2020video, gao2021recent}, where $X^n$ models a set of images known at the encoder. The decoder does not always want to retrieve each image, but has instead a sequence $Y^n$ of particular requests for each image, e.g. detection: cat, dog, car, bike; or scene recognition: street/city/mountain, etc... The encoder does not know the decoder's exact request but has prior information about it (e.g. type of request), which is modeled by $(g(Y_t))_{t \leq n}$.


The problem of Figure \ref{fig:Setting2} relates to the ``restricted inputs'' zero-error problem of Alon and Orlitsky \cite{alon1996source}, as it is obtained as a special case by taking $g$ constant and $f(X,Y) = X$. The optimal rate in ``restricted inputs'' is given by asymptotic chromatic entropies of graph products. Koulgi et al. show in \cite{koulgi2003zero} that this optimal rate is equal to the complementary graph entropy, introduced in \cite{korner1973two} by Körner and Longo. No single-letter expression for these quantities is known. In \cite{marton1993shannon}, Marton shows that these quantities are closely related to the Shannon capacity of a graph (see \cite{shannon1956zero}), which is a wide open problem.

The similar ``unrestricted inputs'' zero-error setting of \cite{alon1996source} has a single-letter formula for the optimal rate, but its zero-error constraint is much stronger than ``restricted inputs'' as $(X^n,Y^n)$ can take values out of the support of $P^n_{X,Y}$.




Now the scheme of Figure 1 has been studied with different coding constraints than zero-error ``restricted inputs'', and the optimal rate has been characterized in each case: the lossless case by Orlitsky and Roche in \cite{orlitsky1995coding}, the lossy case by Yamamoto in \cite{yamamoto1982wyner}, and the zero-error ``unrestricted inputs'' case by Shayevitz in \cite{shayevitz2014distributed}. These results can only be used as bounds here: the zero-error ``restricted inputs'' problem depicted in Figure \ref{fig:Setting2} does not have a characterization of the optimal rate.

Numerous extensions of this problem have been studied recently. The distributed context, for instance, has an additional encoder which encodes $Y$ before transmitting it to the decoder. Achievability schemes have been proposed for this setting by Krithivasan and Pradhan in \cite{krithivasan2011distributed} using abelian groups; by Basu et al. in \cite{basu2020hypergraph} using hypergraphs for the case with maximum distortion criterion; and by Malak and Médard in \cite{malak2020hyper} using hyperplane separations for the continuous lossless case.


Another related context is the network setting, where the function of source random variables from source nodes has to be retrieved at the sink node of a given network. For tree networks, the feasible rate region is characterized by Feizi and Médard in \cite{feizi2014network} for networks of depth one; and by Sefidgaran and Tchamkerten in \cite{sefidgaran2016distributed} under a Markov source distribution hypothesis. In \cite{ravi2018function}, Ravi and Dey consider a bidirectional relay with zero-error ``unrestricted inputs'' and characterize the rate region for a specific class of functions. In \cite{guang2019improved}, Guang et al. study zero-error function computation on acyclic networks with limited capacities, and give an inner bound based on network cut-sets. For both distributed and network settings, the ``restricted inputs'' zero-error problem remains open.

In this paper, we formulate an hypothesis on $P_{X,Y}$ and $g$ that we call ``pairwise shared side information'' that allows us to derive a single-letter characterization of the optimal rate. This hypothesis is satisfied if every pair of source symbols ``share" at least one side information symbol for all output of $g$. It has graph-theoretic interpretations, as the single-letter formula stems from the particular structure of the characteristic graph of the problem: a disjoint union of OR products. Moreover, this result is of practical interest as it covers the cases with $P_{X,Y}$ full-support, without any assumption on $f,g$.


In Section \ref{section:nota}, we present formally the problem of Figure \ref{fig:Setting2}. In Section \ref{section:Gen}, we build the characteristic graphs and give an asymptotic formula for the general case. In Section \ref{section:PSside information}, we make the ``pairwise shared side information'' assumption and give a single-letter characterization of the optimal rate, along with a refinement for $P_{X,Y}$ full-support. We prove it in Section \ref{section:proofmainA} and illustrate it in Section \ref{section:exampleA} with an example.

\section{Problem statement}\label{section:nota}

We denote sequences by $x^n = (x_1, ..., x_n)$. The set of probability distributions over $\mathcal{X}$ is denoted by $\Delta(\mathcal{X})$. The distribution of $X$ is denoted by $P_X \in \Delta(\mathcal{X})$, its support is denoted by $\text{supp } P_X$. Given the sequence length $n\in\N^\star$, we denote by $\Delta_n(\mathcal{X}) \subset \Delta(\mathcal{X})$ the set of empirical distributions of sequences from $\mathcal{X}^n$. We denote by $\lbrace 0,1 \rbrace^*$ the set of binary words. 
The setting of Fig. \ref{fig:Setting2} is described by:
\begin{itemize}[label = -]
\item Four finite sets $\mathcal{X}$, $\mathcal{Y}$, $\mathcal{U}$, $\mathcal{Z}$, a couple of random variables $(X,Y) \in \mathcal{X} \times \mathcal{Y}$ drawn with the distribution $P_{X,Y}$ (with $P_X$ and $P_Y$ full-support), and deterministic functions
    \begin{align}
        & f : \mathcal{X} \times \mathcal{Y} \rightarrow \mathcal{U}, \\
        & g : \mathcal{Y} \rightarrow \mathcal{Z}.
    \end{align}
    For $n$ iterated source uses, we denote by $(X^n, Y^n)$ the sequence of $n$ independent copies of $(X,Y)$, with probability distribution $P^n_{X,Y} \in \Delta(\mathcal{X}^n\times\mathcal{Y}^n)$.
\item The encoder observes the realizations of $X^n, (g(Y_t))_{t \leq n}$ and sends information to the decoder over a noiseless channel of capacity $R\geq0$.
\item The decoder has to recover $(f(X_t, Y_t))_{t \leq n}$ based on the encoder message and the side information $Y^n$.
\end{itemize} 

\begin{definition}[Zero-error source code, achievable rates]\label{}
Given $n\in\mathbb{N}^{\star}$, a $(n, R_n)$-zero-error source code consists of an encoding function $\phi_e : \mathcal{X}^n\times \mathcal{Z}^n \rightarrow \lbrace 0,1 \rbrace^*$ and a decoding function $\phi_d : \mathcal{Y}^n \times \lbrace 0,1 \rbrace^* \rightarrow \mathcal{U}^n$ such that:
\begin{enumerate}
\item the set $\phi_e(\mathcal{X}^n\times \mathcal{Z}^n)$ is prefix-free,
\item $R_n = \frac{1}{n}\mathbb{E}\big[l \circ \phi_e\big(X^n, (g(Y_t))_{t \leq n}\big)\big]$, where $l(\cdot)$ denotes the length of a binary word,
\item the ``restricted inputs'' zero-error property is satisfied: 
    \begin{align}
    & \forall (x^n, y^n) \in \supp P^n_{X,Y}, \nonumber\\
    & \phi_d \Big(y^n, \phi_e\Big(x^n, \big(g(y_t)\big)_{t \leq n}\Big)\Big) = \big(f(x_t, y_t)\big)_{t \leq n}.\label{eq:zeroerr}
\end{align}
\end{enumerate}
A rate $R$ is achievable if there exists a sequence of $(n,R_n)$-zero-error source codes such that $\lim_{n} R_n = R$. The optimal rate is denoted by $R^* = \inf \lbrace R \geq 0 \:|\: R \text{ is achievable} \rbrace$.
\end{definition}

The prefix-free hypothesis guarantees that the decoder knows when the encoder's message stops. A relaxation of this hypothesis is considered in \cite[Theorem 3]{alon1996source}, without influence on the asymptotic optimal rate. Satisfying all three conditions imply a correct decoding with probability one.

\section{General setting}\label{section:Gen}

A probabilistic graph $G$ is a tuple $(\mathcal{V}, \mathcal{E}, P_V)$, where $\mathcal{V}$ is the set of vertices, $\mathcal{E}$ is the set of edges, and $P_V \in \Delta(\mathcal{V})$ is an underlying probability distribution on the vertices. 

We first build the characteristic graph $G_{[n]}$, which is a probabilistic graph that captures the zero-error encoding constraints on a given number $n$ of source uses. It differs from the graphs used in \cite{shayevitz2014distributed}, as we do not need a cartesian representation of these graphs to study the optimal rates. Furthermore, it has a vertex for each possible realization of $\big(X^n, \big(g(Y_t)\big)_{t \leq n}\big)$ known at the encoder, instead of $\mathcal{X}^n$, as in \cite{witsenhausen1976zero} and \cite{alon1996source}. 

\begin{definition}[Characteristic graph $G_{[n]}$]\label{def:G_n}
The characteristic graph $G_{[n]}$ is defined by:
\begin{itemize}[label = -]
\item $\mathcal{X}^n \times \mathcal{Z}^n$ as set of vertices with distribution $P^n_{X,g(Y)}$,
\item $(x^n,z^n)(x'^n,z'^n)$ are adjacent if $z^n = z'^n$ and there exists $y^n \in g^{-1}(z^n)$ such that:
\begin{align}
& \forall t\leq n,\, P_{X,Y}(x_t, y_t) P_{X,Y}(x'_t, y_t) > 0, \label{eq:GnA}\\
\text{and } \, &\exists t \leq n,\,  f(x_t, y_t) \neq f(x'_t, y_t);\label{eq:GnB}
\end{align}
where $g^{-1}(z^n) = \big\lbrace y^n \in \mathcal{Y}^n \:\big|\: \big(g(y_t)\big)_{t \leq n} = z^n \big\rbrace$.
\end{itemize}
\end{definition}

The characteristic graph $G_{[n]}$ is designed with the same core idea as in \cite{witsenhausen1976zero}: $(x^n,z^n)$ and $(x'^n, z^n)$ are adjacent if there exists a side-information symbol $y^n$ compatible with the observation of the encoder (i.e. $z^n = z'^n$ and $y^n \in g^{-1}(z^n)$), such that $f(x^n,y^n) \neq f(x'^n,y^n)$. In order to prevent erroneous decodings, the encoder must map adjacent pairs of sequences to different codewords; hence the use of graph colorings.

\begin{definition}[Coloring, independent subset]
Let $G = (\mathcal{V}, \mathcal{E}, P_V)$ be a probabilistic graph. A subset $\mathcal{S} \subseteq \mathcal{V}$ is independent if $xx' \notin \mathcal{E}$ for all $x,x' \in \mathcal{S}$. Let $\mathcal{C}$ be a finite set (the set of colors), a mapping $c : \mathcal{V} \rightarrow \mathcal{C}$ is a coloring if $c^{-1}(i)$ is an independent subset for all $i \in \mathcal{C}$.
\end{definition}

The chromatic entropy of $G_{[n]}$ gives the best rate of $n$-shot zero-error encoding functions, as in \cite{alon1996source}.

\begin{definition}[Chromatic entropy $H_\chi$]
The chromatic entropy of a probabilistic graph $G = (\mathcal{V}, \mathcal{E}, P_V)$ is defined by
\begin{align}
    H_{\chi}(G) = \inf \big\{ H\big(c(V)\big) \:\big|\: c \text{ is a coloring of } G \big\}.\label{eq:defHchi}
\end{align}
\end{definition}

\begin{theorem}[Optimal rate]\label{th:mainGen}
The optimal rate writes:
\begin{equation}
    R^* = \lim_{n\rightarrow\infty} \frac{1}{n} H_\chi(G_{[n]}).\
\end{equation}
\end{theorem}

\proofoutline{
An encoding function $\phi_e$ is a coloring of $G_{[n]}$ if and only if it satisfies \eqref{eq:zeroerr} with some decoding function $\phi_d$. 

Thus the best achievable rate writes
\begin{align}
R^* = &\: \inf_n \inf_{\phi_e \text{ coloring of } G_{[n]}} H\Big(\phi_e\Big(X^n,\big(g(Y_t)\big)_{t \leq n}\Big)\Big)\\
= & \lim_{n\rightarrow \infty} \frac{1}{n} H_\chi(G_{[n]}).\label{eq:proofthmaingenA}
\end{align}
where \eqref{eq:proofthmaingenA} comes from Fekete's lemma and \eqref{eq:defHchi}.\newline}

A general single-letter expression for $R^*$ is missing, due to the lack of intrinsic structure of $G_{[n]}$. In Section \ref{section:PSside information}, we introduce a hypothesis that gives structure to $G_{[n]}$ and allows us to derive a single-letter expression for $R^*$.

\section{Pairwise shared side information}\label{section:PSside information}

\begin{definition}
The distribution $P_{X,Y}$ and the function $g$ satisfy the ``pairwise shared side information'' condition if
\begin{align}
    \forall z \in \mathcal{Z}, \forall x,x' \in \mathcal{X}, \exists y \in g^{-1}(z), P_{XY}(x,y)P_{XY}(x',y) > 0.\label{eq:HypA}
\end{align}
This means that for all $z$ output of $g$, every pair $(x,x')$ ``shares" at least one side information symbol $y \in g^{-1}(z)$.
\end{definition}

Note that any full-support distribution $P_{X,Y}$ satisfies the ``pairwise shared side information'' hypothesis.

\begin{definition}[AND, OR product]\label{def:ORAND}
Let $G_1\! = \!(\mathcal{V}_1, \mathcal{E}_1, P_{V_1})$, $G_2 = (\mathcal{V}_2, \mathcal{E}_2, P_{V_2})$ be two probabilistic graphs; their AND (resp. OR) product denoted by $G_1 \wedge G_2$ (resp. $G_1 \vee G_2$) is defined by: $\mathcal{V}_1 \times \mathcal{V}_2$ as set of vertices, $P_{V_1}P_{V_2}$ as probability distribution on the vertices, and $(v_1v_2),(v'_1v'_2)$ are adjacent if 
\begin{align}
     v_1v'_1 \in \mathcal{E}_1 \text{ AND }& v_2v'_2 \in \mathcal{E}_2, \\
    \text{resp. }  (v_1v'_1 \in \mathcal{E}_1 \text{ and } v_1 \neq v'_1) \text{ OR }& (v_2v'_2 \in \mathcal{E}_2 \text{ and } v_2 \neq v'_2);\nonumber
\end{align}
with the convention that all vertices are self-adjacent. We denote by $G_1^{\wedge n}$ (resp. $G_1^{\vee n}$) the $n$-th AND (resp. OR) power.\end{definition}

For the ``restricted inputs'' source coding problem in \cite{alon1996source}, the $n$-shot characteristic graph is the $n$-th AND product of the one-shot characteristic graph, and the optimal rate in this problem $\lim_{n \rightarrow \infty} \frac{1}{n} H_{\chi}(G^{\wedge n})$ does not have a single-letter expression. However, for the ``unrestricted inputs'' setting there exists such a formula: the K\"{o}rner graph entropy introduced in \cite{korner1973coding}, which relates to the OR product as shown in Proposition \ref{prop:wheee}. 

\begin{definition}[Körner graph entropy $H_\kappa$]
For all $G = (\mathcal{V}, \mathcal{E}, P_V)$, let $\Gamma(G)$ be the collection of independent sets of vertices in $G$. The K\"{o}rner graph entropy of $G$ is defined by
\begin{align}
    H_\kappa(G) = \min_{V \in W \in \Gamma(G)} I(W;V), \label{eq:wheee}
\end{align}
where the minimum is taken over all distributions $P_{W|V} \in \Delta(\mathcal{W})^{\mathcal{V}}$, with $\mathcal{W} = \Gamma(G)$ and with the constraint that the random vertex $V$ belongs to the random independent set $W$ with probability one, i.e. $V \in W \in \Gamma(G)$ in \eqref{eq:wheee}.
\end{definition}

\begin{proposition}[Properties of $H_\kappa$]\label{prop:wheee}
\cite[Theorem 5]{alon1996source} For all probabilistic graphs $G$ and $G'$,
\begin{align}
    & H_\kappa(G) = \lim_{n \rightarrow \infty} \frac{1}{n} H_{\chi}(G^{\vee n}),\label{eq:Hkappadef}\\
    & H_\kappa(G \vee G') = H_\kappa(G) + H_\kappa(G').\label{eq:Hkappadef2}
\end{align}
\end{proposition}

By using a convex combination of K\"{o}rner graph entropies, we provide a single-letter expression for the optimal rate $R^*$.

\begin{definition}[Auxiliary graph $G^f_z$]
    For all $z \in \mathcal{Z}$, we define the auxiliary graph $G^f_z$ by
    \begin{itemize}[label = -]
        \item $\mathcal{X}$ as set of vertices with distribution $P_{X|g(Y) = z}$,
        \item $xx'$ are adjacent if $f(x,y) \neq f(x',y)$ for some $y \in g^{-1}(z) \cap \supp P_{Y|X = x} \cap \supp P_{Y|X = x'}$.
    \end{itemize}
\end{definition}

\begin{theorem}[Pairwise shared side information]\label{th:mainA}
If $P_{X,Y}$ and $g$ satisfy \eqref{eq:HypA}, the optimal rate writes:
\begin{equation}
    R^* = \sum_{z \in \mathcal{Z}} P_{g(Y)}(z) H_\kappa(G^f_z). \label{eq:Thmaingen}
\end{equation}
\end{theorem}

The proof is in Section \ref{section:proofmainA}, the keypoint is the particular structure of $G_{[n]}$: a disjoint union of OR products.

\begin{remark}
The ``pairwise shared side information'' assumption \eqref{eq:HypA} implies that the adjacency condition \eqref{eq:GnA} is satisfied, which makes $G_{[n]}$ a disjoint union of OR products. Moreover, Körner graph entropies appear in the final expression for $R^*$, even if $G_{[n]}$ is not an $n$-th OR power.
\end{remark}

Now consider the case where $P_{X,Y}$ is full-support. This is a sufficient condition to have \eqref{eq:HypA}. The optimal rate in this setting is derived from Theorem \ref{th:mainA}, which leads to the analytic expression in Theorem \ref{cor:fullsupp}. 

\begin{theorem}[Optimal rate when $P_{X,Y}$ is full-support]\label{cor:fullsupp}
When $P_{X,Y}$ is full-support, the optimal rate writes:
\begin{align}
    R^* = & \: H\big(j(X,g(Y)) \big|g(Y)\big),
\end{align}
where the function $j$ returns a word in $\mathcal{U}^*$, defined by
\begin{align}
    j : & \: \mathcal{X} \times \mathcal{Z} \rightarrow \mathcal{U}^*\\
    & (x,z) \mapsto \big(f(x,y')\big)_{y' \in g^{-1}(z)}.\nonumber
\end{align}
\end{theorem}

\proofoutline{By Theorem \ref{th:mainA}, $R^* = \sum_{z \in \mathcal{Z}} P_{g(Y)}(z) H_\kappa(G^f_z)$. It can be shown that $G^f_z$ is complete multipartite for all $z$ as $P_{X,Y}$ is full support; and it satisfies $H_\kappa(G^f_z) = H\big(j(X, g(Y))\big|g(Y) = z\big)$.}

\section{Example}\label{section:exampleA}

In this example, the ``pairwise shared side information'' assumption is satisfied and $R^*$ is strictly less than a conditional Huffman coding of $X$ knowing $g(Y)$; and also strictly less than the optimal rate without exploiting $g(Y)$ at the encoder.

\begin{figure}[t!]
    \centering
    \begin{tikzpicture}
    \node[] (P) at (-.75,0.25) {$P_{X,Y}$};
    \draw[] (-.75,-.75) -- (-.75,-2.75);
    \draw[] (0,0.25) -- (6.5,0.25);
    \draw[] (0,-.75) -- (6.5,-.75) -- (6.5,-2.75) -- (0,-2.75) -- (0,-.75);
    \draw[dashed] (.25,-.25) -- (3,-.25);
    \draw[dashed] (3.5,-.25) -- (6.25,-.25);
    \node[draw=none] (Y) at (3.25,.5) {$Y$};
    \node[draw=none] (X) at (-1,-1.75) {$X$};
        
    \node[draw=none] (A) at (.5,0) {$0$};
    \node[draw=none] (A) at (1.25,0) {$1$};
    \node[draw=none] (A) at (2,0) {$2$};
    \node[draw=none] (A) at (2.75,0) {$3$};
    \node[draw=none] (A) at (3.75,0) {$4$};
    \node[draw=none] (A) at (4.5,0) {$5$};
    \node[draw=none] (A) at (5.25,0) {$6$};
    \node[draw=none] (A) at (6,0) {$7$};

    \node[draw=none] (A) at (-.5,-1) {$0$};
    \node[draw=none] (A) at (-.5,-1.5) {$1$};
    \node[draw=none] (A) at (-.5,-2) {$2$};
    \node[draw=none] (A) at (-.5,-2.5) {$3$};
    
    \node[draw=none] (A) at (.5,-1) {$0.1$};
    \node[draw=none] (A) at (1.25,-1) {$0.05$};
    \node[draw=none] (A) at (2,-1) {$*$};
    \node[draw=none] (A) at (2.75,-1) {$*$};
    
    \node[draw=none] (A) at (.5,-1.5) {$0.1$};
    \node[draw=none] (A) at (1.25,-1.5) {$*$};
    \node[draw=none] (A) at (2,-1.5) {$0.05$};
    \node[draw=none] (A) at (2.75,-1.5) {$*$};
    
    \node[draw=none] (A) at (.5,-2) {$0.1$};
    \node[draw=none] (A) at (1.25,-2) {$*$};
    \node[draw=none] (A) at (2,-2) {$*$};
    \node[draw=none] (A) at (2.75,-2) {$0.05$};
    
    \node[draw=none] (A) at (.5,-2.5) {$*$};
    \node[draw=none] (A) at (1.25,-2.5) {$0.05$};
    \node[draw=none] (A) at (2,-2.5) {$0.05$};
    \node[draw=none] (A) at (2.75,-2.5) {$0.05$};
    
    \node[draw=none] (A) at (3.75,-1) {$0.05$};
    \node[draw=none] (A) at (4.5,-1) {$0.05$};
    \node[draw=none] (A) at (5.25,-1) {$*$};
    \node[draw=none] (A) at (6,-1) {$*$};
    
    \node[draw=none] (A) at (3.75,-1.5) {$0.05$};
    \node[draw=none] (A) at (4.5,-1.5) {$0.05$};
    \node[draw=none] (A) at (5.25,-1.5) {$0.05$};
    \node[draw=none] (A) at (6,-1.5) {$*$};
    
    \node[draw=none] (A) at (3.75,-2) {$*$};
    \node[draw=none] (A) at (4.5,-2) {$0.05$};
    \node[draw=none] (A) at (5.25,-2) {$*$};
    \node[draw=none] (A) at (6,-2) {$*$};
    
    \node[draw=none] (A) at (3.75,-2.5) {$*$};
    \node[draw=none] (A) at (4.5,-2.5) {$0.05$};
    \node[draw=none] (A) at (5.25,-2.5) {$*$};
    \node[draw=none] (A) at (6,-2.5) {$0.05$};
    
    \node[draw=none] (A) at (1.5,-.5) {$g(Y) = 0$};
    \node[draw=none] (A) at (4.75,-.5) {$g(Y) = 1$};
    \end{tikzpicture}
    
    \begin{tikzpicture}
    \node[] (P) at (-.75,0.25) {$f(\cdot,\cdot)$};
    \node[] (C) at (0,1) {};
    \draw[] (-.75,-.75) -- (-.75,-2.75);
    \draw[] (0,0.25) -- (6.5,0.25);
    \draw[] (0,-.75) -- (6.5,-.75) -- (6.5,-2.75) -- (0,-2.75) -- (0,-.75);
    \draw[dashed] (.25,-.25) -- (3,-.25);
    \draw[dashed] (3.5,-.25) -- (6.25,-.25);
    \node[draw=none] (Y) at (3.25,.5) {$Y$};
    \node[draw=none] (X) at (-1,-1.75) {$X$};
        
    \node[draw=none] (A) at (.5,0) {$0$};
    \node[draw=none] (A) at (1.25,0) {$1$};
    \node[draw=none] (A) at (2,0) {$2$};
    \node[draw=none] (A) at (2.75,0) {$3$};
    \node[draw=none] (A) at (3.75,0) {$4$};
    \node[draw=none] (A) at (4.5,0) {$5$};
    \node[draw=none] (A) at (5.25,0) {$6$};
    \node[draw=none] (A) at (6,0) {$7$};

    \node[draw=none] (A) at (-.5,-1) {$0$};
    \node[draw=none] (A) at (-.5,-1.5) {$1$};
    \node[draw=none] (A) at (-.5,-2) {$2$};
    \node[draw=none] (A) at (-.5,-2.5) {$3$};
    
    \node[draw=none] (A) at (.5,-1) {$a$};
    \node[draw=none] (A) at (1.25,-1) {$b$};
    \node[draw=none] (A) at (2,-1) {$*$};
    \node[draw=none] (A) at (2.75,-1) {$*$};
    
    \node[draw=none] (A) at (.5,-1.5) {$a$};
    \node[draw=none] (A) at (1.25,-1.5) {$*$};
    \node[draw=none] (A) at (2,-1.5) {$b$};
    \node[draw=none] (A) at (2.75,-1.5) {$*$};
    
    \node[draw=none] (A) at (.5,-2) {$b$};
    \node[draw=none] (A) at (1.25,-2) {$*$};
    \node[draw=none] (A) at (2,-2) {$*$};
    \node[draw=none] (A) at (2.75,-2) {$c$};
    
    \node[draw=none] (A) at (.5,-2.5) {$*$};
    \node[draw=none] (A) at (1.25,-2.5) {$c$};
    \node[draw=none] (A) at (2,-2.5) {$c$};
    \node[draw=none] (A) at (2.75,-2.5) {$c$};
    
    \node[draw=none] (A) at (3.75,-1) {$b$};
    \node[draw=none] (A) at (4.5,-1) {$a$};
    \node[draw=none] (A) at (5.25,-1) {$*$};
    \node[draw=none] (A) at (6,-1) {$*$};
    
    \node[draw=none] (A) at (3.75,-1.5) {$a$};
    \node[draw=none] (A) at (4.5,-1.5) {$a$};
    \node[draw=none] (A) at (5.25,-1.5) {$b$};
    \node[draw=none] (A) at (6,-1.5) {$*$};
    
    \node[draw=none] (A) at (3.75,-2) {$*$};
    \node[draw=none] (A) at (4.5,-2) {$b$};
    \node[draw=none] (A) at (5.25,-2) {$*$};
    \node[draw=none] (A) at (6,-2) {$*$};
    
    \node[draw=none] (A) at (3.75,-2.5) {$*$};
    \node[draw=none] (A) at (4.5,-2.5) {$c$};
    \node[draw=none] (A) at (5.25,-2.5) {$*$};
    \node[draw=none] (A) at (6,-2.5) {$c$};
    
    \node[draw=none] (A) at (1.5,-.5) {$g(Y) = 0$};
    \node[draw=none] (A) at (4.75,-.5) {$g(Y) = 1$};
    \end{tikzpicture}
    \caption{An example of $P_{X,Y}$ and $g$ that satisfy \eqref{eq:HypA}; along with the outcomes $f(X,Y)$. The elements outside $\supp P_{X,Y}$ are denoted by $*$.}
    \label{fig:PartSettingA2}
\end{figure}

    Consider the probability distribution and function outcomes depicted in Figure \ref{fig:PartSettingA2}, with $\mathcal{U} = \lbrace a,b,c\rbrace$, $\mathcal{X} = \lbrace 0, ..., 3\rbrace$, $\mathcal{Y} = \lbrace 0, ..., 7\rbrace$, and $\mathcal{Z} = \lbrace 0,1\rbrace$. 
    Let us show that the ``pairwise shared side information'' assumption is satisfied. The source symbols $0,1,2 \in \mathcal{X}$ share the SI symbol $0$ (resp. $5$) when $g(Y) = 0$ (resp. $g(Y) = 1$). The source symbol $3 \in \mathcal{X}$ shares the SI symbols $1,2,3$ with the source symbols $0,1,2$, respectively, when $g(Y) = 0$; and the source symbol $3$ shares the SI symbol $5$ with all other source symbols when $g(Y) = 1$.
    
    Since the ``pairwise shared side information'' assumption is satisfied, we can use Theorem \ref{th:mainA}; the optimal rate writes 
    \begin{align}
        R^* = P_{g(Y)}(0) H_\kappa(G^f_0) + P_{g(Y)}(1) H_\kappa(G^f_1).
    \end{align}
    
    First we need to determine the probabilistic graphs $G^f_0$ and $G^f_1$.
    In $G^f_0$, the vertex $0$ is adjacent to $2$ and $3$, as $f(0,0) \neq f(2,0)$ and $f(0,1) \neq f(3,1)$. The vertex $1$ is also adjacent to $2$ and $3$ as $f(1,0) \neq f(2,0)$ and $f(1,2) \neq f(3,2)$. Furthermore $P_{X|g(Y) = 0}$ is uniform, hence $G^f_0 = (C_4, \text{Unif}(\mathcal{X}))$ where $C_4$ is the cycle graph with $4$ vertices. 
    
    In $G^f_1$, the vertices $1$, $2$, $3$ are pairwise adjacent as $f(1,5)$, $f(2,5)$ and $f(3,5)$ are pairwise different; and $0$ is adjacent to $1$, $2$ and $3$ because of the different function outputs generated by $Y = 4$ and $Y = 5$. Thus, $G^f_1 = (K_4, P_{X|g(Y) = 1})$ with $P_{X|g(Y) = 1} = (\frac{1}{4}, \frac{3}{8}, \frac{1}{8}, \frac{1}{4})$ and $K_4$ is the complete graph with $4$ vertices. An illustration of $C_4$ and $K_4$ is given in Figure \ref{fig:Disjoint}.
    
    Now let us determine $H_\kappa(G^f_0)$ and $H_\kappa(G^f_1)$. On one hand,
    \begin{align}
        H_\kappa(G^f_0) & = H(V_0) - \max_{V_0 \in W \in \Gamma(G^f_0)} H(V_0|W) \label{eq:exampleA10}\\
        & = 2-1 = 1,
    \end{align}
    with $V_0 \sim P_{X|g(Y) = 0} = \text{Unif}(\mathcal{X})$; and where $H(V_0|W)$ in \eqref{eq:exampleA10} is maximized by taking $W = \lbrace 0,1\rbrace$ when $V \in \lbrace 0,1\rbrace$, and $W = \lbrace 2,3\rbrace$ otherwise.
    
    On the other hand, 
    \begin{align}
        H_\kappa(G^f_1) & = \min_{V_1 \in W \in \Gamma(G^f_1)} I(W;V_1) \\
        & = H(V_1) \simeq 1.906,\label{eq:exA2}
    \end{align}
    with $V_1 \sim P_{X|g(Y) = 1}$; where \eqref{eq:exA2} follows from $\Gamma(G^f_1) = \lbrace \lbrace 0\rbrace, ..., \lbrace 3\rbrace\rbrace$, as $G^f_1$ is complete. Hence $R^* \simeq 1.362$.
    
    The rate that we would obtain by transmitting $X$ knowing $g(Y)$ at both encoder and decoder with a conditional Huffman algorithm writes: $R_{\text{Huff}} = H(X|g(Y)) \simeq 1.962$.
    
    The rate that we would obtain without exploiting $g(Y)$ at the encoder is $R_{\text{No } g} = H(X) \simeq 1.985$, because of the different function outputs generated by $Y = 4$ and $Y = 5$.
    
    Finally, $H(f(X,Y)|Y) \simeq 0.875$.
    
    In this example we have 
    \begin{align}
        H(X) = R_{\text{No } g} > R_{\text{Huff}} > R^* > H(f(X,Y)|Y).
    \end{align}
    This illustrates the impact of the side information at the encoder in this setting, as we can observe a large gap between the optimal rate $R^*$ and $R_{\text{No } g}$.

\section{Proof of Theorem \ref{th:mainA}}\label{section:proofmainA}

\subsection{Definitions}
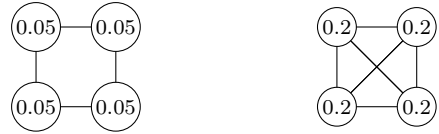
\begin{figure}[t!]
    \centering
    \begin{tikzpicture}
        \foreach \i in {0,1,2,3}{
        \draw[] ($(90*\i-45:0.75)$) -- ($(90*\i + 45:0.75)$);
        \draw[] ($(90*\i-45:0.75)+(4,0)$) -- ($(90*\i + 45:0.75)+(4,0)$);
        \draw[] ($(90*\i-45:0.75)+(4,0)$) -- ($(90*\i + 135:0.75)+(4,0)$);
        }
        \foreach \i in {0,1,2,3}{
        \node[shape = circle, draw = black, inner sep = 1pt, fill = white] (A) at ($(90*\i+45:0.75)$) {\footnotesize$0.05$};
        \node[shape = circle, draw = black, inner sep = 1pt, fill = white] (B) at ($(90*\i+45:0.75) + (4,0)$) {\footnotesize$0.2$};
        }
    \end{tikzpicture}
    \caption{The graph $G$ depicted here with its underlying probability distribution satisfies $G = (C_4, \text{Unif}(\lbrace 1,...,4\rbrace)) \sqcup (K_4, \text{Unif}(\lbrace 1,...,4\rbrace))$; where $C_4$ (resp. $K_4$) is the cycle (resp. complete) graph with $4$ vertices.}
    \label{fig:Disjoint}
\end{figure}

We will use the disjoint union of probabilistic graphs, which generalizes the existing concept of a disjoint union of graphs without underlying probability distribution \cite[Section 1.4]{bondy1976graph}. An example of disjoint union is depicted in Figure \ref{fig:Disjoint}. We also need to formalize the concept of isomorphic probabilistic graphs, i.e. same structure and underlying distribution.

\begin{definition}[Disjoint union of probabilistic graphs]
Let $N \in \mathbb{N}^\star$; let $G = (\mathcal{V}, \mathcal{E}, P_V)$ and for all $i \leq N$, let $\tilde{G}_i = (\mathcal{V}_i, \mathcal{E}_i, P_{V_i})$. We say that $G$ is the disjoint union of the $(\tilde{G}_i)$, denoted by $G = \bigsqcup_{i \leq N} \tilde{G}_i$, if the following is satisfied:
\begin{itemize}[label = - ]
    \item $\mathcal{V}$ is the disjoint union of the sets $(\mathcal{V}_i)_{i \leq N}$, i.e. $\mathcal{V} = \bigcup_i \mathcal{V}_i$ and $\mathcal{V}_i \cap \mathcal{V}_{i'} = \emptyset$ for all $i \neq i'$;
    \item For all $v,v' \in \mathcal{V}$, let $i, i'$ be the unique indexes such that $v \in \mathcal{V}_i$ and $v' \in \mathcal{V}_{i'}$. Then if $i = i'$, $vv' \in \mathcal{E} \Longleftrightarrow vv' \in \mathcal{E}_i$; if $i \neq i'$, $vv' \notin \mathcal{E}$;
    \item For all $i \leq n$, $P_{V|V \in \mathcal{V}_i} = P_{V_i}$.\vspace{-.3cm}
\end{itemize}
\end{definition}

\begin{definition}[Isomorphic probabilistic graphs]
Let $G_1 = (\mathcal{V}_1, \mathcal{E}_1, P_{V_1})$ and $G_2 = (\mathcal{V}_2, \mathcal{E}_2, P_{V_2})$. We say that $G_1$ is isomorphic to $G_2$ (denoted by $G_1 \simeq G_2$) if there exists an isomorphism between them, i.e. a bijection $\psi : \mathcal{V}_1 \rightarrow \mathcal{V}_2$ such that:
\begin{itemize}[label = - ]
    \item For all $v_1, v_1' \in \mathcal{V}_1$, $v_1 v'_1 \in \mathcal{E}_1 \Longleftrightarrow \psi(v_1)\psi(v'_1) \in \mathcal{E}_2$,
    \item For all $v_1 \in \mathcal{V}_1$, $P_{V_1}(v_1) = P_{V_2}\big(\psi(v_1)\big)$.
\end{itemize}
\end{definition}

\subsection{Main proof}
    Let us specify the adjacency condition in $G_{[n]}$ under the assumption \eqref{eq:HypA}. Two vertices are adjacent if they satisfy \eqref{eq:GnA} and \eqref{eq:GnB}; however \eqref{eq:GnA} is always satisfied under \eqref{eq:HypA}. Thus $(x^n,z^n)(x'^n,z^n)$ are adjacent if $z^n = z'^n$ and
    \begin{align}
        \exists y^n \in g^{-1}(z^n), \exists t \leq n, f(x_t, y_t) \neq f(x'_t, y_t).\label{eq:proofAA}
    \end{align}
    It can be observed that the condition \eqref{eq:proofAA} is the adjacency condition of an OR product of adequate graphs; more precisely,
    \begin{align}
        G_{[n]} = \bigsqcup_{z^n \in \mathcal{Z}^n} \bigvee_{t \leq n} G^f_{z_t}.
    \end{align}
    Although $G_{[n]}$ cannot be expressed as an $n$-th OR power, we will show that its chromatic entropy asymptotically coincide with that of an appropriate OR power: we now search for an asymptotic equivalent of $H_\chi(G_{[n]})$.
    
    \begin{definition}
        $\mathcal{S}_n$ is the set of colorings of $G_{[n]}$ that can be written as $(x^n,z^n) \mapsto (T_{z^n}, \tilde{c}(x^n, z^n))$ for some mapping $\tilde{c} : \mathcal{X}^n \times \mathcal{Z}^n \rightarrow \tilde{\mathcal{C}}$; where $T_{z^n}$ denotes the type of $z^n$.\label{def:Sn}
    \end{definition} 
    
    In the following, we define $Z^n \doteq \big(g(Y_t)\big)_{t \leq n}$. Now we need several Lemmas. Lemma \ref{lemma:proofA1} states that the optimal coloring $c(x^n, z^n)$ of $G_{[n]}$ has the type of $z^n$ as a prefix at a negligible rate cost. Lemma \ref{lemma:proofA2} is an adapted version for chromatic entropies of the following observation: minimum colorings on each connected component induce a minimum coloring of the whole graph. Lemma \ref{lemma:proofA3} gives an asymptotic formula for the minimal entropy of the colorings from $\mathcal{S}_n$.
    
    \begin{lemma}\label{lemma:proofA1}
        The following asymptotic comparison holds:
    \begin{align}
        H_\chi(G_{[n]}) = \inf_{\substack{c \text{ coloring of } G_{[n]} \\ \text{s.t. }  c \in \mathcal{S}_n}} H(c(X^n, Z^n)) + O(\log n).
    \end{align}
    \end{lemma}
    \begin{lemma}\label{lemma:proofA2}
        Let $N \in \mathbb{N}^\star$; let $G = (\mathcal{V}, \mathcal{E}, P_V)$ and for all $i \leq N$, let $\tilde{G}_i = (\mathcal{V}_i, \mathcal{E}_i, P_{V_i})$ be probabilistic graphs such that $G = \bigsqcup_i \tilde{G}_i$ and $\tilde{G}_1 \simeq ... \simeq \tilde{G}_N$. Then we have $H_\chi(G) = H_\chi(\tilde{G}_1)$.
    \end{lemma}
    \begin{lemma}\label{lemma:proofA3}
    The following asymptotic comparison holds:
    \begin{align}
        \inf_{\substack{c \text{ coloring of } G_{[n]} \\ \text{s.t. }  c \in \mathcal{S}_n}} H(c(X^n, Z^n)) =  n \sum_{z \in \mathcal{Z}} P_{g(Y)}(z) H_\kappa(G^f_z) + o(n).
    \end{align}
    \end{lemma}
    
    The keypoint of the proof of Lemma \ref{lemma:proofA1} is the asymptotically negligible entropy of the prefix $T_{Z^n}$ of the colorings of $\mathcal{S}_n$. 
    
    Lemma \ref{lemma:proofA2} is proved using the concavity of the entropy, which implies the following: an optimal coloring colors all the isomorphic connected components the same way. 
    
    The proof of Lemma \ref{lemma:proofA3} relies on the decomposition $G_{[n]} = \bigsqcup_{Q_n \in \Delta_n(\mathcal{Z})} G_{[n]}^{Q_{n}}$, where $G_{[n]}^{Q_{n}}$ is the subgraph induced by the vertices $(x^n, z^n)$ such that the type of $z^n$ is $Q_n$. We show that $G_{[n]}^{Q_{n}}$ is a disjoint union of isomorphic graphs whose chromatic entropy is given by Lemma \ref{lemma:proofA2} and \eqref{eq:Hkappadef2}: $\big|H_\chi(G^{Q_n}_{[n]}) - n \sum_{z \in \mathcal{Z}} Q_n (z) H_\kappa(G^f_z)\big| \leq n\epsilon_n$. Finally, uniform convergence arguments enable us to conclude.
    
    Now let us combine these results together:
    \begin{align}
        R^* & = \frac{1}{n} H_\chi(G_{[n]}) + o(1) \label{eq:proofA50}\\
        & = \frac{1}{n} \inf_{\substack{c \text{ coloring of } G_{[n]} \\ \text{s.t. }  c \in \mathcal{S}_n}} H(c(X^n, Z^n)) + o(1) \label{eq:proofA51}\\
        & = \sum_{z \in \mathcal{Z}} P_{g(Y)}(z) H_\kappa(G^f_z) + o(1),\label{eq:proofA52}
    \end{align}
    where \eqref{eq:proofA50} comes from Theorem \ref{th:mainGen}, \eqref{eq:proofA51} comes from Lemma \ref{lemma:proofA1}, and \eqref{eq:proofA52} comes from Lemma \ref{lemma:proofA3}. The proof of Theorem \ref{th:mainA} is complete.

\subsection{Proof of Lemmas \ref{lemma:proofA1}, \ref{lemma:proofA2}, \ref{lemma:proofA3}}

\textbf{Proof of Lemma \ref{lemma:proofA1}.}
    Let $c_n^*$ be the coloring of $G_{[n]}$ with minimal entropy. Then we have:
    \begin{align}
        H_\chi(G_{[n]}) & = \inf_{c \text{ coloring of } G_{[n]}} H(c(X^n, Z^n))\label{eq:proofA1} \\
        & \leq \inf_{\substack{c \text{ coloring of } G_{[n]} \\ \text{s.t. }  c \in \mathcal{S}_n}} H(c(X^n, Z^n))\label{eq:proofA2} \\
        & = \inf_{\substack{c : (x^n,z^n) \\ \mapsto (T_{z^n}, \tilde{c}(x^n, z^n))}} H(T_{Z^n}, \tilde{c}(X^n, Z^n))\label{eq:proofA3} \\
        & \leq H(T_{Z^n}) + H(c_n^*(X^n, Z^n)) \label{eq:proofA4} \\
        & = H_\chi(G_{[n]}) + O(\log n), \label{eq:proofA5}
    \end{align}
    where \eqref{eq:proofA3} comes from Definition \ref{def:Sn}; \eqref{eq:proofA4} comes from the subadditivity of the entropy, and the fact that $(x^n, z^n) \mapsto (T_{z^n}, c^*_n (x^n, z^n))$ is a coloring of $G_{[n]}$ that belongs to $\mathcal{S}_n$; and \eqref{eq:proofA5} comes from  $H(T_{Z^n}) = O(\log n)$, as $\log|\Delta_n(\mathcal{Z})| = O(\log n)$. The desired equality comes from the bounds $H_\chi(G_{[n]})$ and $H_\chi(G_{[n]}) + O(\log n)$ on \eqref{eq:proofA2}.\newline
    
    \textbf{Proof of Lemma \ref{lemma:proofA2}.} Let $(\tilde{G}_i)_{i \leq N}$ be isomorphic probabilistic graphs and $G$ such that $G = \bigsqcup_i \tilde{G}_i$. Let $c_1^* : \mathcal{V}_1 \rightarrow \mathcal{C}$ be the coloring of $\tilde{G}_1$ with minimal entropy, and let $c^*$ be the coloring of $G$ defined by
    \begin{align}
        c^* : \: & \mathcal{V} \rightarrow \mathcal{C} \\
        & v \mapsto c_1^* \circ \psi_{i_v \rightarrow 1}(v),
    \end{align}
    where $i_v$ is the unique integer such that $v \in \mathcal{V}_{i_v}$, and $\psi_{i_v \rightarrow 1} : \mathcal{V}_{i_v} \rightarrow \mathcal{V}_1$ is an isomorphism between $\tilde{G}_{i_v}$ and $\tilde{G}_1$. In other words $c^*$ applies the same coloring pattern $c^*_1$ on each connected component of $G$. We have
    \begin{align}
    H_\chi(G) & \leq H(c^*(V)) \label{eq:proofA01}\\
    & = h\Big(\sum_{j \leq N} P_{i_V}(j) P_{c^*(V_j)}\Big) \\
    & = h\Big(\sum_{j \leq N} P_{i_V}(j) P_{c_1^*(V_1)}\Big) \label{eq:proofA03}\\
    & = H(c^*_1(V_1)) \\
    & = H_\chi(\tilde{G}_1),\label{eq:proofA05}
    \end{align}
    where $h$ denotes the entropy of a distribution; \eqref{eq:proofA03} comes from the definition of $c^*$; and \eqref{eq:proofA05} comes from the definition of $c_1^*$.
    
    Now let us prove the upper bound on $H_\chi(\tilde{G}_1)$. Let $c$ be a coloring of $G$, and let $i^* \doteq \argmin_i H(c(V_i))$ (i.e. $i^*$ is the index of the connected component for which the entropy of the coloring induced by $c$ is minimal). We have
    \begin{align}
    H(c(V)) & = h\Big(\sum_{j \leq N} P_{i_V}(j) P_{c(V_j)}\Big) \label{eq:proofA10} \\
    & \geq \sum_{j \leq N} P_{i_V}(j) h(P_{c(V_j)}) \label{eq:proofA11} \\
    & \geq \sum_{j \leq N} P_{i_V}(j) H(c(V_{i^*})) \label{eq:proofA12} \\
    & \geq H_\chi(\tilde{G}_{i^*}),\label{eq:proofA13} \\
    & = H_\chi(\tilde{G}_1),\label{eq:proofA14}
    \end{align}
    where \eqref{eq:proofA11} follows from the concavity of $h$; \eqref{eq:proofA12} follows from the definition of $i^*$; \eqref{eq:proofA13} comes from the fact that $c$ induces a coloring of $\tilde{G}_{i^*}$; \eqref{eq:proofA14} comes from the fact that $\tilde{G}_1$ and $\tilde{G}_{i^*}$ are isomorphic. Now, we can combine the bounds \eqref{eq:proofA05} and \eqref{eq:proofA14}: for all coloring $c$ of $G$ we have
    \begin{align}
        H_\chi(G) \leq H_\chi(\tilde{G}_1) \leq H(c(V)),
    \end{align}
    which yields the desired equality when taking the infimum over $c$.\newline

    \textbf{Proof of Lemma \ref{lemma:proofA3}.}
        For all $Q_n \in \Delta_n(\mathcal{Z})$, let 
        \begin{align}
            G^{Q_n}_{[n]} = \bigsqcup_{\substack{z^n \in \mathcal{Z}^n \\ T_{z^n} = Q_n}} \bigvee_{t \leq n} G^f_{z_t},
        \end{align}
        with the probability distribution induced by $P^n_{X,Z}$. This graph is formed of the connected components of $G_{[n]}$ whose corresponding $z^n$ has type $Q_n$. We need to find an equivalent for $H_\chi(G^{Q_n}_{[n]})$. Since $G^{Q_n}_{[n]}$ is a disjoint union of isomorphic graphs, we can use Lemma \ref{lemma:proofA2}:
        \begin{align}
            & H_\chi(G^{Q_n}_{[n]}) = H_\chi\bigg(\bigvee_{z \in \mathcal{Z}} (G^f_z)^{\vee n Q_n(z)}\bigg). \label{eq:proofA35}
        \end{align}
        On one hand, 
        \begin{align}
            H_\chi\bigg(\bigvee_{z \in \mathcal{Z}} (G^f_z)^{\vee n Q_n(z)}\bigg) & \geq H_\kappa\bigg(\bigvee_{z \in \mathcal{Z}} (G^f_z)^{\vee n Q_n(z)}\bigg) \label{eq:proofA36}\\
            & = n \sum_{z \in \mathcal{Z}} Q_n (z) H_\kappa(G^f_z), \label{eq:proofA37}
        \end{align}
        where \eqref{eq:proofA36} comes from $H_\kappa \leq H_\chi$ \cite[Lemma 14]{alon1996source}, \eqref{eq:proofA37} comes from \eqref{eq:Hkappadef2}. On the other hand, 
        \begin{align}
            \!\!\!H_\chi\bigg(\bigvee_{z \in \mathcal{Z}} (G^f_z)^{\vee n Q_n(z)}\bigg) & \leq \sum_{z \in \mathcal{Z}} Q_n(z) H_\chi((G^f_z)^{\vee n}) \label{eq:proofA38}\\
            & = n \sum_{z \in \mathcal{Z}} Q_n (z) H_\kappa(G^f_z) + n\epsilon_n, \label{eq:proofA39}
        \end{align}
        where $\epsilon_n \doteq \max_z \frac{1}{n} H_\chi((G^f_z)^{\vee n}) - H_\kappa(G^f_z)$ is a quantity that does not depend on $Q_n$ and satisfies $\lim_{n \rightarrow \infty} \epsilon_n = 0$; \eqref{eq:proofA38} comes from the subadditivity of $H_\chi$. Combining equations \eqref{eq:proofA35}, \eqref{eq:proofA37} and \eqref{eq:proofA39} yields
        \begin{align}
             \left|H_\chi(G^{Q_n}_{[n]}) - n \sum_{z \in \mathcal{Z}} Q_n (z) H_\kappa(G^f_z)\right| \leq n\epsilon_n. \label{eq:proofA39b}
        \end{align}
        
        Now, we have an equivalent for $H_\chi(G^{Q_n}_{[n]})$.
        \begin{align}
            & \inf_{\substack{c \text{ coloring of } G_{[n]} \\ \text{s.t. }  c \in \mathcal{S}_n}} H(c(X^n, Z^n)) \label{eq:proofA30} \\
            = & \inf_{\substack{c : (x^n,z^n) \\ \mapsto (T_{z^n}, \tilde{c}(x^n, z^n))}} H(\tilde{c}(X^n, Z^n)|T_{Z^n}) + H(T_{Z^n})\label{eq:proofA31} \\
            = & \inf_{\substack{c : (x^n,z^n) \\ \mapsto (T_{z^n}, \tilde{c}(x^n, z^n))}} \sum_{Q_n \in \Delta_n(\mathcal{Z})} P_{T_{Z^n}}(Q_n)\nonumber \\
            & H(\tilde{c}(X^n, Z^n)| T_{Z^n} = Q_n) + O(\log n) \label{eq:proofA32} \\
            = & \sum_{Q_n \in \Delta_n(\mathcal{Z})} P_{T_{Z^n}}(Q_n) \inf_{c_{Q_n} \text{ coloring of } G^{Q_n}_{[n]}} \nonumber\\
            & H(c_{Q_n}(X^n, Z^n)| T_{Z^n} = Q_n) + O(\log n)\label{eq:proofA33} \\
            = & \sum_{Q_n \in \Delta_n(\mathcal{Z})} P_{T_{Z^n}}(Q_n) H_\chi(G^{Q_n}_{[n]}) + O(\log n)\label{eq:proofA34} \\
            = \: & \sum_{Q_n \in \Delta_n(\mathcal{Z})} P_{T_{Z^n}}(Q_n) \left(n\sum_{z \in \mathcal{Z}} Q_n (z) H_\kappa(G^f_z) \pm n\epsilon_n\right) \nonumber\\
            & + O(\log n) \label{eq:proofA40}\\
            = \: & n \sum_{Q_n \in \Delta_n(\mathcal{Z})} 2^{-nD(Q_n\|P_{g(Y)}) + o(n)} \left(\sum_{z \in \mathcal{Z}} Q_n (z) H_\kappa(G^f_z)\right) \nonumber\\
            & \pm n \epsilon_n + O(\log n) \label{eq:proofA41} \\
            = \: & n \sum_{z \in \mathcal{Z}} P_{g(Y)}(z) H_\kappa(G^f_z) + o(n), \label{eq:proofA42} 
        \end{align}
        where \eqref{eq:proofA32} comes from $H(T_{Z^n}) = O(\log n)$, as $\log|\Delta_n(\mathcal{Z})| = O(\log n)$; \eqref{eq:proofA33} follows from the fact that the entropy of $\tilde{c}$ can be minimized independently on each $G^{Q_n}_{[n]}$; \eqref{eq:proofA34} follows from the definition of $G^{Q_n}_{[n]}$; \eqref{eq:proofA40} comes from \eqref{eq:proofA39b}; \eqref{eq:proofA41} comes from \cite[Lemma 2.6]{csiszar2011information} and the fact that $\epsilon_n$ does not depend on $Q_n$.

\bibliographystyle{IEEEtran}

\end{document}